\documentclass[12pt]{article}
\usepackage{amsfonts,amsmath,amssymb}

\usepackage{bigints}
\usepackage[colorinlistoftodos]{todonotes}

\usepackage{lscape,slashed}
\usepackage{multirow}
\usepackage{color}
\usepackage{bm}
\usepackage{comment}
\usepackage{amssymb,amsmath,yhmath,mathrsfs,graphicx}
\usepackage{amstext}
\usepackage{amscd}
\usepackage{graphics}
\usepackage{epsfig}
\usepackage{shadow}
\usepackage[all]{xy}
\usepackage{bbm}
\usepackage{nicefrac}
\definecolor{blue}{rgb}{.255,.41,.884} 
\definecolor{red}{rgb}{1, 0, 0} 
\definecolor{green}{rgb}{.196,.804,.196} 
\definecolor{yellow}{rgb}{1,.648,0} 
\definecolor{pink}{rgb}{1,0.5,0.5}
\usepackage{lscape,slashed}
\usepackage{multirow}
\usepackage{color}
\usepackage{bm}
\usepackage{comment}
\usepackage{graphics}
\usepackage{epsfig}
\usepackage{shadow}
\usepackage[all]{xy}
\usepackage{hyperref}
\usepackage{bbm}
\usepackage{nicefrac}
\def\r{\rho}
\def\l{\lambda}
\def\s{\sigma}

\def\d{\partial}
\def\del{\delta}
\def\m{\mu}
\def\n{\nu}

\def\a{\alpha}
\def\b{\beta}

\def\nn{\nonumber}
\def\be{\begin{equation}}
\def\ee{\end{equation}}
\def\bea{\begin{eqnarray}}
\def\eea{\end{eqnarray}}
\def\de{\partial}
\def\ft#1#2{{\textstyle{\frac{\scriptstyle #1}{\scriptstyle #2} } }}

\begin{document}
\begin{flushright} BRX TH-6641 \\
CALT-TH 2018-046
\end{flushright}

\vspace{.8cm}
\setcounter{footnote}{0}
\begin{center}
{\Large{\bf
Are all identically conserved geometric tensors metric variations of actions? A status report}
    }\\[10mm]

{\sc S.\ Deser$^{\star}$ and
Y.\ Pang$^\ddag$
\\[6mm]}

{\em\small
$^\star$California Institute of Technology, Pasadena, CA 91125 and\\
Brandeis University, Waltham,
MA 02454,
USA\\
 {\tt deser@brandeis.edu}}\\[5mm]
{\em\small
$^\ddag$
 Mathematical Institute, University of Oxford, \\
	Woodstock Road, Oxford OX2 6GG, U.K.
\\ {\tt Yi.Pang@maths.ox.ac.uk}}\\[5mm]

\bigskip

\bigskip

{\sc Abstract}\\
\end{center}

{\small
\begin{quote}
Noether's theorem, that local gauge variations of gauge invariant actions are identically conserved (more tautologically, that gauge variations of gauge invariants vanish) was established a century ago. Its converse, in the geometric context: are all identically conserved local symmetric tensors variations of some coordinate invariant action? remains unsolved to this day. We survey its present state and discuss some of our concrete attempts at a solution, including a significant improvement. For notational simplicity, details are primarily given in $D=2$, but we discuss generic $D$ as well.




\end{quote}
}

\newpage

\section{Introduction}
 Noether's theorem is a textbook truism that the field equations of gauge theories--Maxwell, Yang-Mills, Einstein {\it et.al}--obey conservation,``Bianchi'', identities as a consequence of their Lagrangian origins: The actions being invariant, their local gauge variations vanish. But the latter are just the
divergences of the action's field variations: It suffices for models to be Lagrangian for them to
obey gauge identities. But is it also necessary--are all identically conserved currents derived from actions? This converse hypothesis is almost as old as Noether's and remains unsolved--for the gravitational case--despite its simple form and intuitive appeal. Over the last few decades, only limited success has been achieved. For instance, when the tensor has at most two metric derivatives, $\partial^2 g$, it is Lagrangian \cite{L1}; at $\partial^3g$ order, \cite{H} proved the Lagrangian
nature in definite signature spaces. That assumption was lifted in \cite{L2} in $D=3$, while \cite{A1} gave the general $\partial^3 g$ proof in all $D$. For a detailed history, see, e.g., \cite{A2}. To our knowledge, there is no result beyond $\partial^3 g$ until our present $\partial^6g$ one. This is not merely a formal conjecture, but has direct physical consequences: Non-Lagrangian terms have recently been proposed as alternative geometrical models. But the physics requires them to be separately conserved: Since coordinate invariant matter actions' stress-tensors are identically conserved (on matter shell), irrespective of their couplings, if any, to gravity, the proposed field equations,
\be
G_{\m\n}(g) + E_{\m\n}(g) =T_{\m\n} ({\rm matt}; g)
\ee
imply that the non-Lagrangian gravitational addition $E_{\m\n}$ must be identically conserved, since both the Lagrangian gravity part $G_{\m\n}$ (including $G_{\m\n}=0$ ) and--as we saw\footnote{A recent suggestion \cite{Bekenstein} that a matter Lagrangian is not needed to specify matter systems, but only conservation of the stress-tensors, can be understood in this light as being entirely equivalent to the standard lore:  A correct stress tensor is always the metric variation of an action, and is conserved IFF the matter field equations are invoked.} --$T_{\m\n}({\rm matt})$ both are. Hence counterexamples to the necessity hypothesis, if they existed, would be of physical interest and conversely their absence would remove a sea of models.
We shall first review the vector gauge theories, where there are manifold counter-examples to the conjecture, before coming to the gravitational story. Concentrating on the most elementary geometrical systems, those in $D=2$ where only the scalar curvature enters, we will discuss some
 differential and integral approaches to exhibit the nature of some of the obstacles involved as well as all-order versus perturbative attempts; in the former case we have succeeded in reaching several derivative order improvements over past results. Higher-dimensional  similarities and differences will also be discussed. For completeness, we emphasize that we are only interested in local currents constructed only from metric and its local derivatives, as these are meant to be primary models.
There are of course non-local quantum contributions to effective actions (from anomalies inter alia) but these are all action-generated anyway. We also underline the irrelevance of (action-defined) matter's $T_{\m\n}$ details. The complementary question of non-action matter sources' gravity problems is treated in \cite{d1}.
Given the simplicity and plausibility of the hypothesis, we cannot help but feel some obvious proof is being overlooked; perhaps
this r\'{e}sum\'{e} will attract one!

\section{Vectors}
A  sufficiently general set of field equations, first in the abelian, $D=4$ Maxwell, case, is
\be
M^{\n}=\de_{\m}\Big[X(F^2, \widetilde{F}F)F^{\m\n}\Big]=0\,,
\label{em}
\ee
where $\widetilde{F}^{\m\n}$ is the ($D=4$) dual of $F_{\m\n}$ and we have used only its two simplest, algebraic, invariants in the arbitrary function $X$.
The divergence identities $\de_\n M^\n=0$ are manifest from the antisymmetry of $F$ contracted with the symmetric $\de_\m
\de_\n$, irrespective of $X$. However, not all such $M$ are $A_\m$ variation of a Lagrangian: they
must obey the usual Helmholz integrability conditions, which set stringent limits on the $X$.
So here identical conservation does NOT require an action. Perhaps surprisingly this is not some purely linear, abelian property, but holds also for non-abelian fields: there, we replace $\de_\m$ by the usual covariant color derivatives $D_\m$ whose commutator is now the non-abelian field strength, $[D_\m\,,D_\n]\sim F_{\m\n}$. Yet the generalization of \eqref{em} remains transverse owing to the antisymmetry of the structure constants:  $f_{abc}F^{b\m\n}F^c_{\m\n}=0$ (the arguments of $X$ are now the  (color-singlet) traces of $F^2$ and $ \widetilde{F}F$). Again, only the algebraic factor: antisymmetry, is relevant.
\section{Gravity}
We now come to our problem:  the origin of identically conserved geometric tensors. The formalism
is enormously simplified by working first in $D=2$, where all essentials are already present, index
proliferation is at a minimum and the issues are manifest. Only the scalar curvature $R$ and its covariant derivatives, $\nabla^n R$, (since $R_{\m\n}=\ft12g_{\m\n}R$), and explicit metrics contracting indices are present.
Our convention is
\be
R=g^{\m\n}R_{\m\n}=g^{\m\n}R^{\alpha}{}_{\m\alpha\n}=g^{\m\n}\left(\partial_{\a}\Gamma^{\a}{}_{\m\n}-\partial_{\n}\Gamma^{\a}{}_{\a\m}+\Gamma^{\a}{}_{\a\b} \Gamma^{\b}{}_{\m\n}-\Gamma^{\a}{}_{\n\b} \Gamma^{\b}{}_{\m\a}\right)\,.
\ee
Its variation is
\be
\frac{\delta R(x)}{\delta g^{\m\n}(y)}=\left[\frac12 g_{\m\n} R + (g_{\m\n} \nabla^2-\nabla_\m\nabla_\n)\right]\delta^{(2)}(x-y).
\label{Rvar}
\ee
Note that the $\nabla\nabla$ part of $\delta R$ is the covariant version of the flat space transverse projector
$O^{\m\n} = [\eta^{\m\n} \partial^2 -\partial^\m \partial^\n]$, but it is of course no longer transverse; there are none in curved space. Indeed this is the 2D version of the flat superpotentials
$V^{\m\n}= \partial_{\alpha} \partial_{\beta} H^{[\m\alpha][\n\beta]}$, where $H$ has the algebraic symmetries of the Riemann tensor, so $V$ is identically conserved. In $D=2$, $H$ degenerates into $\varepsilon^{\m\alpha} \varepsilon^{\n\beta} S$ where $S$ is a scalar, namely into the $O^{\m\n}$ above. First, a reminder of why invariant action-based tensors are conserved here (non-invariant actions' variations are of course not even tensors). The variation of
\be
A= \int d^2x L \left(g_{\m\n}; \nabla^n R\right),~n\ge 0
\ee
is
\be
\frac{\delta A}{\delta g_{\m\n}(x)}\Big\vert_{\rm total}=  \frac{\delta A}{\delta g_{\m\n}(x)}\Big\vert_{R\,{\rm const}}+\int d^2y\frac{\delta R(y)}{\delta g_{\m\n}(x)}\frac{\delta A}{\delta R(y)}\Big\vert_{g\,{\rm const} } \,,
\label{Avar}
\ee
and of course the Noether identity $\nabla_\n \frac{\del A}{\del g_{\m\n}}\Big\vert_{\rm total}=0$ holds because $A$ is invariant under arbitrary coordinate variations, $\del g_{\m\n}= \nabla_{(\m}\xi_{\n)}$. Note that both terms in \eqref{Avar} are ``normal'' tensors, as against
``projector'' ones, $O^{\m\n} S$--this point is critical to our problem, so we explain it. (Ex-)projectors are of course tensors, but strange ones whose divergences are NOT in general total derivatives: despite the notation, $\nabla_\n(O^{\m\n} S)$ is of the form $S\d R$ (or $R\d S$, depending on choice); that
is manifestly NOT always the divergence of any regular, NON-$OS$, tensor--for example if $S=(\d R)^2$. The Lagrangian case is the one where $OS$ is normal, because it also can be written as $\delta R/\delta g$, so for
$S=\frac{\del A}{\sqrt{-g}\del R}\vert_g$ we recover \eqref{Avar}.

The above illustrates sufficiency;
Now for necessity: are there NON-Lagrangian identically conserved $X^{\m\n}( g_{\m\n};\nabla^n R)$? In the vector cases, we saw that such (vector) terms existed because one merely algebraically contracted antisymmetric with symmetric indices, unlike the differential nature of the present problem. The lowest-level cases are easy: if $X^{\m\n}$ is $R$-independent, it must be proportional to $g^{\m\n}$, namely to a cosmological action $L= \sqrt{-g}$. Likewise, $X=X(g;R)$ obviously comes from an
$L=\sqrt{-g} f(R)$. This is no longer so obvious when $X$ does depend on derivatives of  $R$. We must fall back on the projector basis of flat space conservation for inspiration. As we saw above, if the $R$-dependence is such that a scalar $S$ is of the form $\frac{\del A}{\sqrt{-g}\del R}\vert_g$, then  $\int d^2y\sqrt{-g} \delta R(y)/\delta g_{\m\n}(x) S(y)$ is the $R$-variation of an action and the total conserved current is its sum with $\frac{\del A}{\sqrt{-g}\del g_{\m\n}(x)}\vert_R$. The inspiration is of course $\eqref{Rvar}$, showing that the flat $O^{\m\n}$ must be extended to the curved one, plus the (natural) $gR$-term.
We can now state the general problem in its tersest form, at least in the present approach.
Are there NON-Lagrangian solutions of the local equation $\nabla_\n(O^{\m\n}S +Z^{\m\n})=0$, where $Z$ is a ``normal'' tensor, $S$ a scalar and $O$ the $\delta R/\delta g_{\m\n}$ of \eqref{Rvar}?
So far the only way a compensating ``normal'' $Z$ can exist is for $OS$ to have a normal divergence as discussed above. Although we have not
succeeded in settling the question, it seems so intuitively simple that these lines may inspire a resolution.
In higher $D$, there are a few novel wrinkles, such as the existence of 4-index
$O^{\m\n\r\l}$ from the variations of the--identically conserved--Einstein tensor, multiplied by a 2-tensor $S_{\r\l}$ and of course the complications of dependence on the index-rich (covariant derivatives of)  Ricci and Riemann tensors. These are all examples of the general superpotential $\partial_{\alpha} \partial_{\beta} H^{[\m\alpha][\n\beta]}$ mentioned earlier. Then there are Chern-Simons like operators in odd $D$, and finally for $D>4$ the Lanczos-Lovelock \cite{L1,L3}\footnote{\cite{L3} merely noted the quadratic curvature topological invariants in $D=4$, namely Gauss-Bonnet and its axial counterpart $\int d^4x\widetilde{R}R$, while \cite{L1} showed that the G-B action becomes dynamical for $D>4$ and listed all such extensions.} actions' variations have no contributions from their curvature dependence, but rather entirely from their explicit metric dependence, in complete contrast with $D=2$, where the latter is trivial.

Let us now look (back in $D=2$) at the problem, first in a perturbative way. [For space reasons, we will be very terse about the still unsolved approaches.]. We seek an identically conserved tensor $X_{\m\n}$
\be
X_{\mu\nu}=(\nabla_{\mu}\nabla_{\nu}-g_{\mu\nu}\Box)S-\ft12SRg_{\mu\nu}+Z_{\mu\nu}\,,
\label{xdec}
\ee
whose vanishing divergence means that
\be
\nabla^{\mu}Z_{\mu\nu}=\ft12S\partial_{\nu}R\,,
\label{Zeq1}
\ee
an equation that resembles that of a scalar-tensor model with $R$ an independent scalar. In a weak field expansion about flat space,
\be
g_{\m\n}\simeq \eta_{\m\n}+\epsilon h_{\m\n},\quad \epsilon \ll 1\,,
\label{weak}
\ee
the leading term in \eqref{Zeq1} becomes
\be
\partial^{\m}Z^{(L)}_{\m\n}=\ft12S^{(L)}\partial_{\nu}R^{(L)}\,,
\label{Zeq2}
\ee
in an obvious notation; all covariant derivatives are here replaced by partials. The right hand side of \eqref{Zeq2} is annihilated by Euler-Lagrangian operator (since it kills all total derivatives), is a necessary, but not sufficient,  condition for conservation. While we have not been able to solve the resulting condition iteratively in general, we have at least succeeded in pushing the known results \cite{A1} several orders higher in derivatives of $R$, namely to $\partial^6$ in the metric,  as we briefly sketch. When $X^{\m\n}$ depends  only on the first six derivatives of the metric, we can in fact construct a non-perturbative proof by solving \eqref{Zeq1} directly. In this case, $S$ depends on at most the second derivative of the $R$ so the most general $Z_{\m\n}$ must take the form
\be
Z_{\m\n}=A(R,T)(\nabla_\m\nabla_\n-g_{\m\n}\Box)R+B(R,T)\nabla_\m R\nabla_\n R+g_{\m\n}C(R,T),\quad T\equiv (\partial R)^2\,.
\ee
Notice that the $\nabla^2R$ terms in $Z_{\m\n}$ must appear in the combination $O_{\m\n}R$ or $\nabla^{\m}Z_{\m\n}$ would depend on $\nabla^3 R$. Now demanding 
\be
\nabla^\m Z_{\m\n}=\ft12S(R,\nabla R,\nabla^2 R)\partial_\n R
\ee
yields 
\be
B=-\frac{\partial A}{\partial R}-2\frac{\partial C}{\partial T}\,.
\label{6thsol}
\ee
In deriving \eqref{6thsol}, we have used the $D=2$ identities for any scalar quantity $\phi$
\bea
2\phi_{\m\n}\phi^{\n\l}\nabla_\l\phi-2\phi_{\m\n}\nabla^\n\phi\Box\phi&=&\nabla_\m\phi[(\Box\phi)^2-\phi_{\n\l}\phi^{\n\l}],~ \phi_{\m\n}\equiv\nabla_\m\nabla_\n\phi\,,\nn
\\ [0.08cm]
g_{\m\n}\phi_{\l\rho}\nabla^{\l}\phi\nabla^\rho\phi-2\phi_{\l(\m}\nabla_{\n)}\phi\nabla^\l\phi &=&(\partial\phi)^2(g_{\m\n}\Box-\nabla_\m\nabla_\n)\phi-\Box\phi\nabla_\m\phi\nabla_\n\phi\,.
\eea
One can show that the particular $Z_{\m\n}$ satisfying \eqref{6thsol} results from varying the following action with respect to $g^{\m\n}$ with $R$ fixed
\be
A=\int d^2x\sqrt{g} \left(A\nabla_{\m}\log T\nabla^\m R-2C\right)\,.
\ee
Therefore, general covariance implies $\sqrt{g}S=\frac{\delta A}{\delta R}\vert_{g }$ which contains at most second derivative of $R$. By the reasoning given in previous paragraphs, $X_{\m\n}$ constructed in \eqref{xdec} comprises the general divergence free symmetric tensor depending on at most $\partial^6g$. This procedure can of course continue to higher orders with the encounter of complicated new Schouten-type identities at each order. We did not proceed further because our main goal is an all-order proof which seems to be beyond the limit of the current approach. However, equation \eqref{Zeq1} does provide a link between scalar-tensor models and the divergenceless symmetric $D=2$ tensor, the latter being four derivative orders higher than the former in their respective fundamental fields.

A different approach to the problem would be  to establish that $X^{\m\n}$ obeys the integrability condition
\be
\frac{\delta \sqrt{g}X^{\m\n}(x)}{\delta g_{\r\s}(y)}=\frac{\delta \sqrt{g}X^{\r\s}(y)}{\delta g_{\m\n}(x)}\,,
\label{int0}
\ee
namely, $\frac{\delta \sqrt{g}X^{\m\n}(x)}{\delta g_{\r\s}(y)}$ is a formally self-adjoint differential operator comprised of the Riemann tensor and its covariant derivatives. The integral form of \eqref{int0} can be expressed as
\be
\int_M\big(\delta_2(\sqrt{g} X^{\mu\nu}) \delta_1 g_{\mu\nu}-\delta_1(\sqrt{g} X^{\mu\nu}) \delta_2g_{\mu\nu}\big)=0\, ,
\label{intg1}
\ee
for arbitrary variations $ \delta_1 g$ and $ \delta_2 g$. To approach our goal \eqref{intg1},
first define the functional
\be
A_X(Y):=\int_M\sqrt{g} X^{\mu\nu} Y_{\mu\nu}\, ,
\ee
in which the tensor $Y_{\m\n}$ has finite support on $M$. Conservation of $X^{\mu\nu}$ implies this functional vanishes when $Y$ is the
Lie derivative of the metric with respect to a
compactly supported vector field:
\be
A_X({\mathcal  L}_\xi g):=2\int_M \sqrt{g}\,  X^{\mu\nu} \nabla_\mu \xi_\nu=0\, .
\ee
Here ${\mathcal L}$ denotes the Lie derivative and we have used that
${\mathcal L}_\xi g_{\mu\nu} =\nabla_\mu \xi_\nu + \nabla_\nu \xi_\mu$.
Hence, the  variation of $A({\mathcal  L}_\xi g)$ also vanishes so that
\be
\delta_1 A_X({\mathcal  L}_\xi g)= \int_M \big(\delta_1( \sqrt{g} X^{\mu\nu}) {\mathcal L}_\xi g_{\mu\nu} + \sqrt{g} X^{\mu\nu} {\mathcal L}_\xi \delta_1 g_{\mu\nu}\big)=0\, .
\ee
The functional $A_X(Y)$ is diffeomorphism-invariant, so a variation  $\delta_2 A_X(\delta_1 g)$ with $\delta_2 \delta_1 g= {\mathcal  L}_\xi (\delta_1 g)$ also vanishes. This gives
\be
\int_M\big(\delta_2(\sqrt{g} X^{\mu\nu}) \delta_1 g_{\mu\nu}+
\sqrt{g} X^{\mu\nu} {\mathcal  L}_\xi (\delta_1 g)\big)
=0\, .
\ee
The difference of the above two displays
\be
\int_M\big(\delta_2(\sqrt{g} X^{\mu\nu}) \delta_1 g_{\mu\nu}-\delta_1(\sqrt{g} X^{\mu\nu}) \delta_2g_{\mu\nu}\big)\big|_{\delta_2 g={\mathcal  L}_\xi g}=0\, .
\label{intg3}
\ee
Were $\delta_2 g$ not restricted to variations of the form ${\mathcal  L}_\xi g$, this would complete the proof.
However, $X^{\m\n}$ has three components in $D=2$, of which only two are affected  by \eqref{intg3}. We have unfortunately been unable to complete this ``integral" approach either.

\section{Comments}
We have reviewed and summarized the current standing of a century-old conjecture-validity of
the converse of Noether's theorem:  are all identically conserved geometrical 2-tensors the metric variations of some invariant action? This intuitively attractive proposition has proved remarkably recalcitrant to date, although we have managed to push the proof to sixth derivative order in the metric.  A number of quite different approaches have been pursued and we have summarized them by concentrating on the simplest curved space dimension, $D=2$, where the problem is most clearly stated without the obscuring higher $D$ index proliferation.  A proof (or indeed disproof) in $D=2$ all but guarantees the same for all $D$.
There are important physical consequences of this seemingly formal question to real physics:
Of the many attempts to go beyond GR, addition of non-Lagrangian terms on the ``left hand side"
of the field equations requires them to be identically conserved, since both $G_{\m\n}$ and  the (Lagrangian-based) matter stress tensors on their mass shell are. This would close the floodgates
to a wide range of speculation. [Conversely, in the unlikely event that there are such tensors, a whole new field would open up!]  In string theory, one always obtains $DX$=0 equation for the target space fields from the world-sheet BRST invariance. So if our conjecture is true,
it also implies that all stringy gravity models are Lagrangian. 

We have used locality as a physical demand. If that is lifted, it is trivial to provide counter-examples, albeit non-symmetric ones, such as $X^{\m\n}=(\nabla^\m\Box^{-1}\nabla^\n-g^{\m\n})S$ (conserved on one index). Finally, we have not investigated the recently proposed \cite{bhmrt,op} amusing $D=3$ models whose $X$-divergences only vanish on-shell.

\section*{Acknowledgements}
We thank Andrew Waldron for his useful suggestions in the early stages of this endless saga. The work of S.D. is supported by the U.S. Department of Energy, Office of Science, Office of High Energy Physics, under Award Number DE-SC0011632. The work of Y.P. is supported by
a Newton International Fellowship of the UK Royal Society.


\begin{thebibliography}{20}

\bibitem{L1}
  D.~Lovelock,
  ``The Einstein tensor and its generalizations,''
  J.\ Math.\ Phys.\  {\bf 12} (1971) 498.

\bibitem{H}
 G.~W.~Horndeski, 
 {``Divergence-free third order tensorial concomitants of a pseudo-Riemannian metric"}, 
 Tensor {\bf 29} (1975), 21–29.
 
 \bibitem{L2}
D.~Lovelock, 
{``Divergence-free third order concomitants of the metric tensor in three
dimensions"}, Topics in differential geometry (in memory of Evan Tom Davies), pp.
87-98. Academic Press, New York, 1976.


\bibitem{A1}
 Ian~M.~Anderson, Juha, Pohjanpelto,
{``Variational principles for natural divergence-free tensors in metric
field theories,''}
 Journal of Geometry and Physics {\bf 62} (2012) 2376-2388.


\bibitem{A2}
Ian~M.~Anderson, {`` The variational bicomplex''}, Academic Press, Boston 1994.



\bibitem{Bekenstein}
  J.~D.~Bekenstein and B.~R.~Majhi,
  {``Is the principle of least action a must?,''}
  Nucl.\ Phys.\ B {\bf 892} (2015) 337.
 
\bibitem{d1}
 S.~Deser,
{``Non-Lagrangian Gauge Field Models are Physically Excluded,''}, Phys.\ Lett.\ B {\bf 790} (2019) 408.

\bibitem{L3}
C.~Lanczos,
``A remarkable property of the Riemann-Christoffel tensor in four dimensions,"
Annals Math. {\bf 39} (1938) 842.






\bibitem{bhmrt}
  E.~Bergshoeff, O.~Hohm, W.~Merbis, A.~J.~Routh and P.~K.~Townsend,
  ``Minimal Massive 3D Gravity,''
  Class.\ Quant.\ Grav.\  {\bf 31} (2014) 145008.

\bibitem{op}
  M.~Ozkan, Y.~Pang and P.~K.~Townsend,
  {``Exotic Massive 3D Gravity,''} JHEP {\bf 1808} (2018) 035.

\end{thebibliography}
\end{document}